\documentclass{PoS}
\usepackage{amsmath}
\usepackage{subfigure}
\DeclareMathOperator{\Di}{D}
\DeclareMathOperator{\Pf}{Pf}
\DeclareMathOperator{\Tr}{{\rm Tr}}
\DeclareMathOperator{\sign}{sign}
\newcommand{\gino}{\lambda}
\newcommand{\bgino}{\bar{\lambda}}
\newcommand{\erw}[1]{\langle #1\rangle}
\newcommand{\I}{\ensuremath{\mathrm{i}}}
\newcommand{\arxiv}[1]{[arXiv:\href{http://arxiv.org/abs/#1}{#1}]}
\newcommand{\aetap}{\text{a--}\eta'}
\newcommand{\api}{\text{a--}\pi}
\newcommand{\afn}{\text{a--}f_0}
\title{The gluino-glue particle and relevant scales for the simulations
of supersymmetric Yang-Mills theory}

\ShortTitle{SUSY Yang-Mills theory}

\author{\speaker{Georg Bergner}%
\\
University of Frankfurt, Institute for Theoretical Physics\\
E-mail: \email{g.bergner@physik.uni-frankfurt.de}}
\author{Istvan Montvay\\
Deutsches Elektronen-Synchrotron DESY,\\
Notkestr. 85, D-22603 Hamburg, Germany}
\author{Gernot M\"unster, Dirk Sandbrink, Umut D.\ \"Ozugurel\\
University of M\"unster, Institute for Theoretical Physics\\
Wilhelm-Klemm-Strasse 9, D-48149 M\"unster, Germany}

\abstract{%
Supersymmetric Yang-Mills theory is in several respects different from QCD
and pure Yang-Mills theory. Therefore, a reinvestigation of the scales, at
which finite size effects and lattice artifacts become relevant, is
necessary. Both, finite size effects and lattice artifacts, induce a
breaking of supersymmetry. In view of the unexpected mass gap between
bosonic and fermionic particles an estimation of these effects is essential.
}

\FullConference{The 30th International Symposium on Lattice Field Theory\\
June 24 - 29,  2012\\
Cairns, Australia}
\begin{document}
\section{Introduction}

Supersymmetry plays a central role in theoretical models for elementary
particle physics beyond the Standard Model. Therefore it is important to
gain knowledge about the properties of supersymmetric theories. Much of what
is known about supersymmetric models is based on tree-level considerations
or comes from perturbation theory. However, various important
characteristics, like the existence of mass-degenerate supermultiplets of
particles, are of a non-perturbative nature. Therefore it is desirable to
study them by means of non-perturbative methods.

The simplest supersymmetric model including gauge fields is the
supersymmetric Yang-Mills theory (SYM). It describes interacting gluons and
their supersymmetric partners, the gluinos, which are Majorana fermions in
the adjoint representation of the gauge group SU($N_{c}$). The (on-shell)
Lagrangian of SYM in Minkowski space is
\begin{equation}
\mathcal{L}=\Tr\left[-\frac{1}{4}
F_{\mu\nu}F^{\mu\nu}+\frac{\I}{2}\bar{\lambda}\gamma^\mu
D_\mu\lambda{-\frac{m_g}{2}\bar{\lambda}\lambda} \right ] \,,
\end{equation}
where $\lambda$ is the gluino field, $A_{\mu}$ the gluon field, 
$F_{\mu\nu}$ the non-Abelian field strength, and 
\begin{equation}
D_\mu \lambda^{a} = \partial_{\mu} \lambda^{a} +
g\,f_{abc} A^{b}_{\mu} \lambda^{c}
\end{equation}
denotes the gauge covariant derivative in the adjoint representation. The
gluino mass term breaks supersymmetry softly.

SYM is similar to QCD in some respect \cite{Amati:1988ft}. It is
asymptotically free and is assumed to show confinement. The ``physical''
particles are bound states of gluons and gluinos, and if supersymmetry is
unbroken, they would form supermultiplets. The non-perturbative properties
one would like to investigate with the lattice simulations include: (1) the
spontaneous breaking of chiral symmetry, $Z_{2 N_{c}} \to Z_{2}$, that
manifests itself in the non-vanishing vacuum expectation value
$\langle\lambda\lambda\rangle \neq 0$, (2)~the confinement of static quarks,
indicated by a linear rise in the static quark potential, which is an
evidence for the confining nature of the theory, (3) the spectrum of bound
states, that can be compared to the predictions of low energy effective
actions, (4) whether supersymmetry is spontaneously broken, (5) the
restoration of SUSY in the continuum limit of a lattice regularisation.

The particle content of SYM is expected to consist of colour neutral bound
states of gluons and gluinos, forming supermultiplets. Based on effective
Lagrangeans, it has been predicted \cite{Veneziano:1982ah,Farrar:1997fn}
that the low-lying particles form two chiral supermultiplets, each
consisting of a scalar, a pseudoscalar, and a fermionic spin 1/2 particle.
One of them contains a $0^-$gluinoball ($\aetap \ \sim \
\overline{\lambda}\gamma_5\lambda$), a $0^+$gluinoball ($\afn \ \sim \
\overline{\lambda} \lambda$), and a spin 1/2 gluino-glueball ($\chi \ \sim
\sigma^{\mu\nu}\, \Tr(F_{\mu\nu}\lambda)$), the other one a $0^-$ glueball,
a $0^+$ glueball, and a gluino-glueball. Both supermultiplets contain an
exotic particle state called gluino-glueball, which is a spin 1/2 Majorana
fermion. Such a bound state containing a single fermion does not occur in
QCD, but analogous particles exist in models similar to QCD with an
arbitrary number of quark flavours in the adjoint representation.

We study SYM non-perturbatively in the framework of regularisation on a
lattice. Supersymmetry is generically broken in any non-trivial theory on
the lattice \cite{Bergner:2009vg}. The particle spectrum and the
supersymmetric Ward identities can show how it is restored in the continuum
limit or whether there is a possible remnant breaking of SUSY by
non-perturbative effects. The existence of supermultiplets is an important
signal for the supersymmetric limit of the theory.

Previous work by our collaboration on SYM on the lattice, concerning the
nonperturbative items mentioned above, is reported in
\cite{Demmouche:2010sf,Bergner:2011wf} and references given there. The
results obtained there have not yet shown the expected degeneracy of the
fermionic and bosonic masses. The mass of the gluino-glueball appeared to be
larger than the other masses of its lightest possible superpartners.
However, the masses were obtained at a fixed lattice spacing and without a
detailed analysis of the finite size effects. Below we shall discuss our
recent calculations, which indicate that the influence of the finite lattice
spacing is larger than expected and provides a possible source of the
supersymmetry breaking in the simulation.

%%%%%%%%%%%%%%%%%%%%%%%%%%%%%%%%%%%%%%%%%%%%%%%%%%%%%%%%%%%%%%%
\section{SUSY on the Lattice}

In our work we employ the formulation of SYM on a lattice by Curci and
Veneziano \cite{Curci:1986sm}, where the gluinos are described by Wilson
fermions in the adjoint representation. SYM has also been investigated with
domain wall fermions \cite{Fleming:2000fa,Giedt:2008xm,Endres:2009yp} and
overlap fermions \cite{Kim:2011fw}, which, however, require significantly
more computing resources than Wilson fermions for large lattice volumes and
small lattice spacings.

The lattice action of SYM in our setup is
\begin{equation}
  S_L=\beta \sum_p\left(1-\frac{1}{N_c}\mbox{Re}\,\Tr U_p\right)
   +\frac{1}{2}\sum_{xy} \bar{\lambda}_x(\Di)_{xy}\lambda_y\, ,
\end{equation}
where $\Di$ is the Wilson-Dirac operator
\begin{eqnarray}
 (\Di)_{x,a,\alpha;y,b,\beta}
    &=&\delta_{xy}\delta_{a,b}\delta_{\alpha,\beta}\nonumber\\
    &&-\kappa\sum_{\mu=1}^{4}
      \left[(1-\gamma_\mu)_{\alpha,\beta}(V_\mu(x))_{ab}
                          \delta_{x+\mu,y}
      +(1+\gamma_\mu)_{\alpha,\beta}(V^\dag_\mu(x-\mu))_{ab}
                          \delta_{x-\mu,y}\right],
\end{eqnarray}
and $(V_\mu(x))_{ab}$ are the gauge link variables in the adjoint
representation. We are currently considering the gauge group SU(2) with
generators $T^a = \tau_a / 2$, in which case
\begin{equation}
(V_\mu(x))_{ab} = 2\,\textrm{Tr}\,(U_{x\mu}^\dagger T_a U_{x\mu} T_b).
\end{equation}
The hopping parameter $\kappa$ is related to the bare gluino mass via
$\kappa=1/(2m_g+8)$.

In order to reduce lattice artifacts, in our simulations we actually use the
tree-level Symanzik improved gauge action, and one level of stout smearing
applied to the link fields in the Wilson-Dirac operator.

From considerations of chiral and SUSY Ward identities it is expected
\cite{Curci:1986sm} that a fine-tuning of the bare gluino mass parameter
(i.e.\ $\kappa$) in the continuum limit is sufficient to approach the chiral
symmetry and supersymmetry of the continuum theory. This tuning is most
efficiently done by means of the mass of the unphysical adjoint pion
($\api$). This particle is the pion in the corresponding theory with two
Majorana fermions in the adjoint representation. The correlator of this
particle is the connected contribution of the $\aetap$ correlator. The
$\api$ is not a physical particle in SYM. However, it can be defined in a
partially quenched setup, in the same way as for one-flavour QCD
\cite{Farchioni:2007dw}. On the basis of arguments involving the
OZI-approximation of SYM \cite{Veneziano:1982ah}, the adjoint pion mass is
expected to vanish for a massless gluino. The corresponding value of
$\kappa_c$ is most easily obtained from the dependence of the $\api$-mass on
$\kappa$.

The numerical simulations are done using a polynomial hybrid Monte Carlo
(PHMC) algorithm with a two step polynomial approximation and reweighting
\cite{Montvay:2005tj,Demmouche:2010sf}. The functional integral over
Majorana fermions yields the Pfaffian
$\Pf(C\Di)=\sign(\Pf(C\Di))\sqrt{\det(C\Di)}$ of the antisymmetric fermion
matrix multiplied by the charge conjugation matrix $C$. The PHMC implements
the positive weight $\sqrt{\det(\Di)}$, while the sign $\sign(\Pf(C\Di))$ is
taken into account by reweighting. The reweighting factors are obtained from
the number of negative real eigenvalues of $\Di$.

%%%%%%%%%%%%%%%%%%%%%%%%%%%%%%%%%%%%%%%%%%%%%%%%%%%%%%%%%%%%%%%
\section{The particle spectrum on the lattice}

The most interesting particles are the possible candidates for the lowest
lying multiplets. These multiplets consist of a fermionic, a scalar, and a
pseudoscalar particle. On the lattice a mixing between possible operators
representing the corresponding quantum numbers is expected. As
representations of the bosonic scalar and pseudoscalar particles we have
measured glueballs and gluinoballs. The fermionic particle is represented by
the gluino-glueball.

The correlation function of the $0^{++}$ glueball is calculated by smeared
plaquette-plaquette correlations. Variational smearing techniques (APE and
HYP smearing) are used to get a reasonable signal for the particle mass.
Unfortunately the signal for the $0^{-+}$ glueball turned out to be
insufficient for a determination of the mass with our current statistics.

The gluinoballs are similar to flavour singlet mesonic operators in QCD.
Hence they contain disconnected contributions and require the computation of
all-to-all propagators. In a graphical representation of the correlators the
disconnected contributions show up as two disconnected fermion loops, e.~g.\
for the $\aetap$
\begin{equation}
\erw{\bgino(x)\gamma_5 \gino(x)\, \bgino(y)\gamma_5 \gino(y)}
=\erw{\;\raisebox{-0.2cm}{\includegraphics[width=1cm]{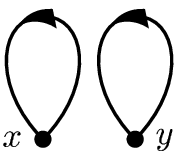}}
-2\;\raisebox{-0.2cm}{\includegraphics[width=1cm]{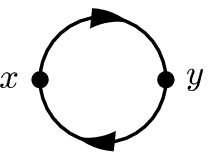}}\;}.
\end{equation}
From the computation of flavour singlet mesons in QCD it is well known that
the all-to-all propagators introduce additional statistical noise in the
computation of an observable. We observe this effect in our simulations when
we compare the larger statistical errors of the $\aetap$ with the small
error of the $\api$. We have implemented and tested several techniques to
determine the all-to-all propagators \cite{Farchioni:2004ej}. In our recent
simulations a combination of the truncated eigenmode approximation and the
stochastic estimator method has been used. Our truncated eigenmode
approximation is implemented in such a way that from the eigenvalues
reweighting factors can be obtained to improve the polynomial approximation
of the PHMC algorithm. The stochastic estimator method was improved with
iterations of a truncated solver, see \cite{Bali:2009hu} for the details of
these methods. With these techniques the $\aetap$ mass can be obtained with
a reasonable precision, while the determination of the $\afn$ mass is still
not satisfactory.

\begin{figure}
\subfigure[$0^+$ glueball]{
\includegraphics[width=7.0cm]{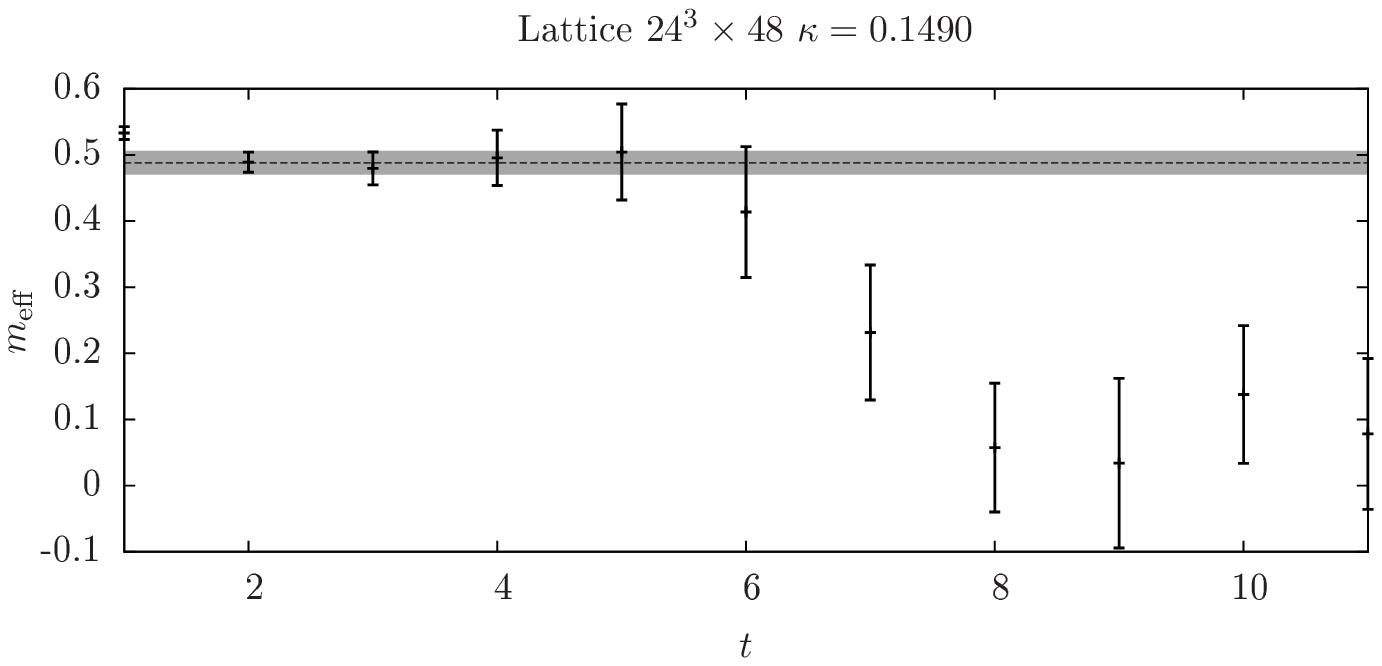} 
}
\subfigure[$\aetap$]{
\includegraphics[width=7.0cm]{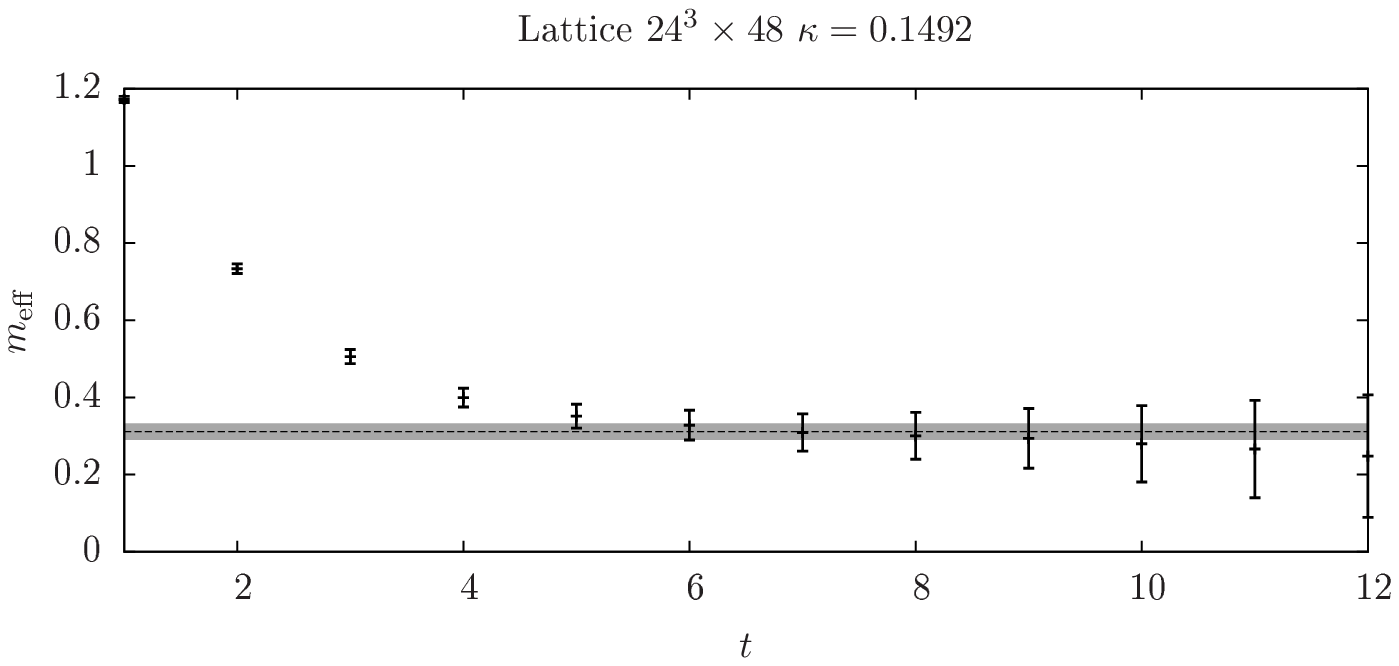}
}
\subfigure[$\afn$]{
\includegraphics[width=7.0cm]{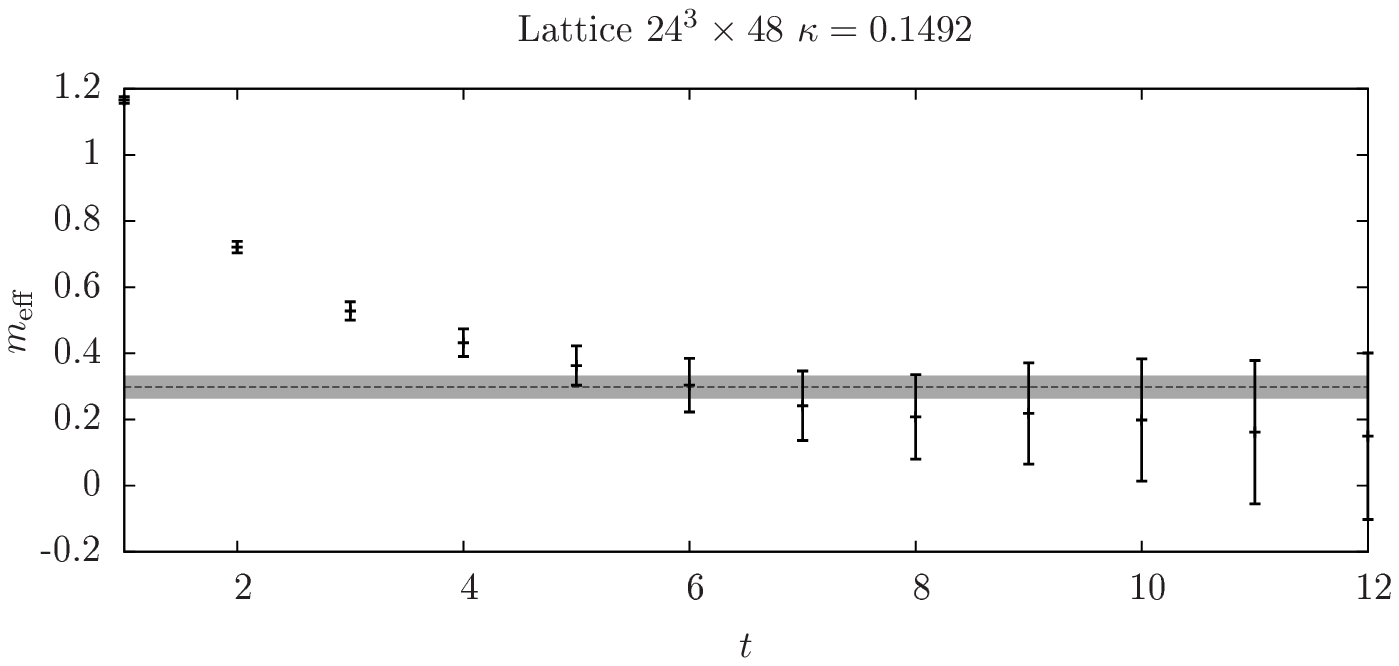}
}
\subfigure[gluino-glue]{
\includegraphics[width=7.0cm]{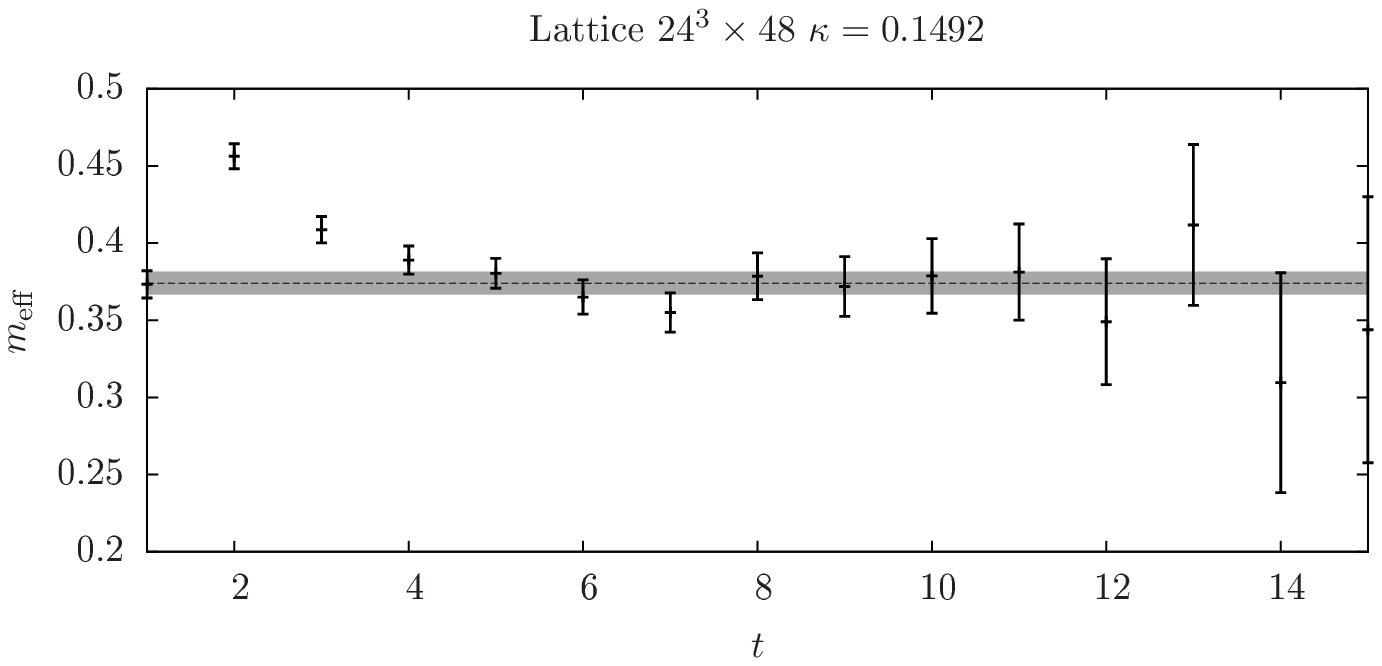}
}
\subfigure[$\api$]{
\includegraphics[width=7.0cm]{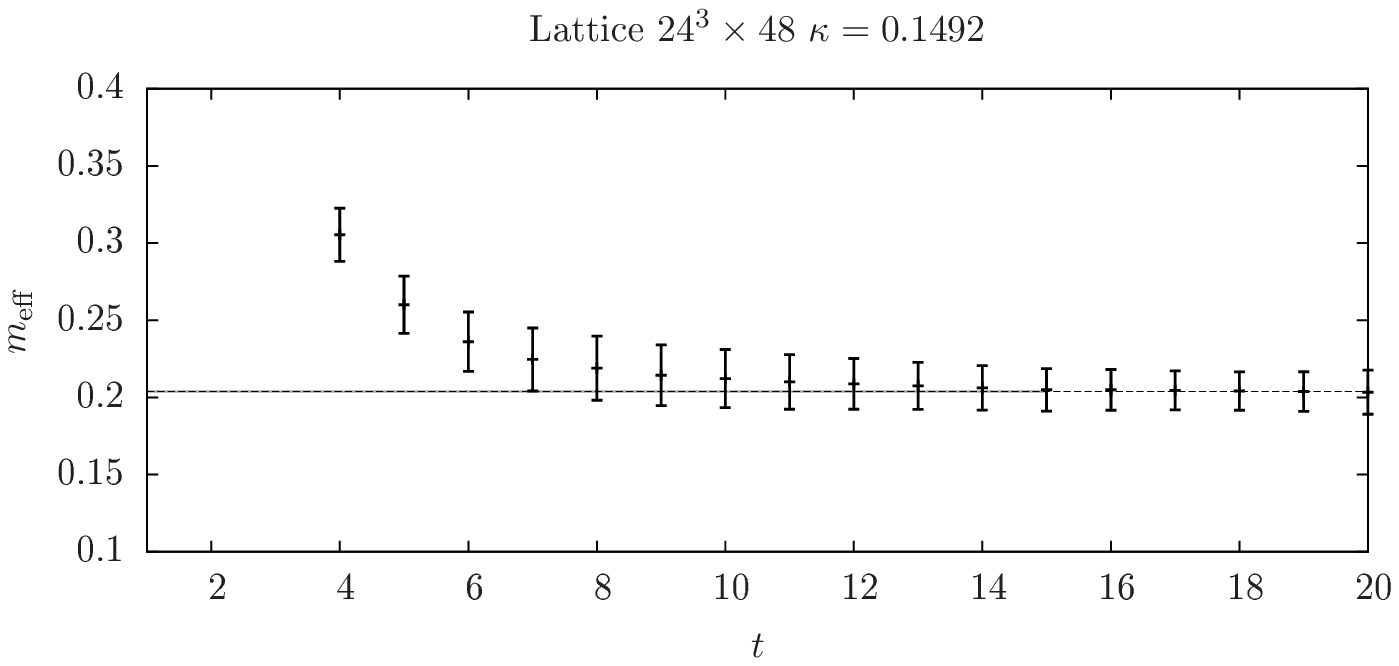}
}
\caption{Examples for the effective masses of the particles measured in
supersymmetric Yang-Mills theory. Time slice distances $t,t+a$ are used to
compute the effective mass $m_{\text{eff}}$ at distance $t$ assuming a
single $\cosh$ ($\sinh$) behavior. The mass and $t$ are given in units of
the lattice spacing $a$. The lattice size is $24^3\times 48$. The parameters
of the simulations are $\beta=1.75$ and $\kappa=0.1490/0.1492$.}
\label{fig:effmass}
\end{figure}

In contrast to the difficulties in the bosonic sector, the mass of the
fermionic gluino-glueball can been obtained most reliably. The corresponding
lattice operator is $\sigma^{\mu\nu}\Tr [F_{\mu\nu}\gino]$, where the field
strength on the lattice is represented by the clover plaquette. A
combination of APE (applied on $F_{\mu\nu}$) and Jacobi smearing (applied on
$\gino$) leads to a further reduction of the noise for this signal. Plots of
the effective masses for this particles can be found in
Fig.~\ref{fig:effmass}.

%%%%%%%%%%%%%%%%%%%%%%%%%%%%%%%%%%%%%%%%%%%%%%%%%%%%%%%%%%%%%%%
\section{Finite size effects and lattice artifacts}

In previous work of our collaboration a rather large gap between fermionic
and bosonic masses has been observed \cite{Demmouche:2010sf}. These
simulations were done at $\beta=1.6$ and lattice sizes up to $24^3\times
48$. It would, however, be premature to draw conclusions from this about
possible supersymmetry breakings in continuum SYM. There are several limits
involved in the determination of physical results for this model, namely the
infinite volume limit, the continuum limit, and the chiral limit, which have
to be taken in this order. With present resources it is not possible to
carry out this program. In recent simulations we have, however, been able to
obtain some estimates for the relevance and influence of these limits.

\begin{figure}
\begin{center}
\includegraphics[width=9cm]{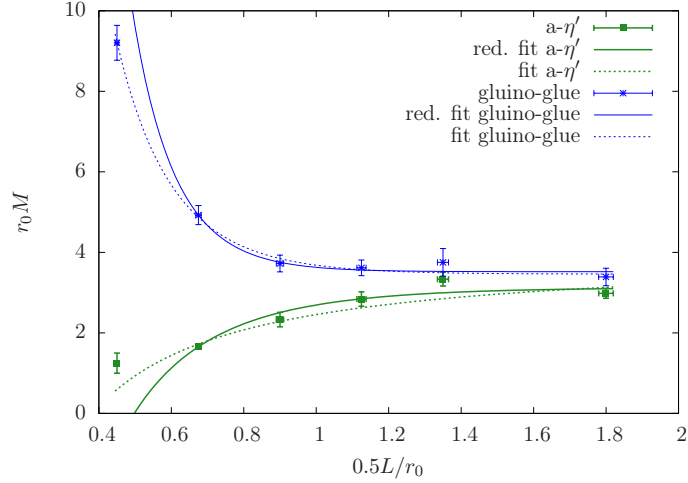} 
\caption{The influence of the finite box size $L$ on the mass gap between
the bosonic $\aetap$ and the fermionic gluino-glueball. These results were
obtained at a fixed value of the bare parameter $\kappa=0.1490$
($\beta=1.75$). The mass is, as indicated, shown in units of $r_0^{-1}$, and
$L$ in units of $r_0/0.5$, where $r_0$ is the Sommer parameter. When $r_0$
is set to its QCD value, the units of $L$ would correspond to $\text{fm}$. }
\label{fig:Lmassgap}
\end{center}
\end{figure}

We have performed a detailed investigation of the finite size effects in
simulations of $8^3\times 16$, $12^3\times 24$, $16^3\times 36$, $20^3\times
40$, $24^3\times 48$, and $32^3\times 64$ lattices at $\beta=1.75$
\cite{Bergner:2012rv}. For the asymptotic dependence of masses on the box
size $L$ we have taken \cite{Luscher:1985dn,Munster:1984zf}
\begin{equation}
m(L) \approx m_0 + C  L^{-1} \exp \left( -\alpha m_0 L\right)
\end{equation}
for large values of $L$. Results have been obtained at fixed bare parameter
$\kappa$ and at a fixed mass of $\api$. As explained above, the best signal
is obtained for the masses of the gluino-glueball and the $\aetap$. The
difference between their masses can be taken as an estimate of the mass gap
between the particles of a supermultiplet. The dependence of the mass gap on
the finite box size is shown in Fig.~\ref{fig:Lmassgap}.

The mass gap is clearly increased by the influence of the finite volume. The
supersymmetry breaking terms, that are present due to the lattice
discretisation are hence increased by the finite size of the lattice.
However, this effect becomes rather small already at moderately large
lattice sizes. Therefore, finite size effects cannot be the source of the
observed SUSY-breaking effects.

\begin{figure}
\begin{center}
\includegraphics[width=9cm]{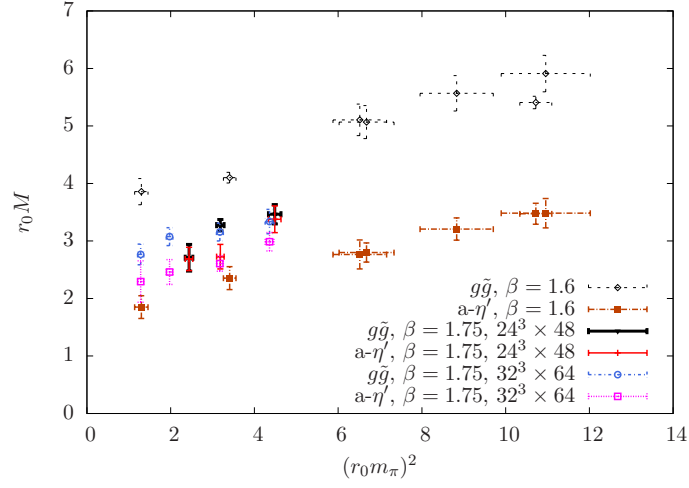} 
\caption{The mass of the $\aetap$ and the gluino-glueball at two different
$\beta$, where the larger value corresponds to the smaller lattice spacing.
The masses are shown as a function of the squared mass of the adjoint pion
($m_\pi$). The results from the two different lattice sizes $24^3\times 48$
and $32^3\times 64$ are found to be consistent. All masses are given in
units of the Sommer parameter $r_0$.}
\label{fig::latspac}
\end{center}
\end{figure}

We have also been able to estimate the influence of the finite lattice
spacing from simulations at a second larger value of $\beta$. The difference
between the results at $\beta=1.6$ ($a\approx 0.08\ \text{fm}$) and
$\beta=1.75$ ($a\approx 0.06\ \text{fm}$) are shown in
Fig.~\ref{fig::latspac}.
They indicate that indeed the lattice artifacts are the most relevant source
for a bias in the spectrum of the particles.

Possible alternative ways to reduce the lattice artifacts are currently
being explored. For example, an increased smearing in the fermionic part of
the action seems to reduce their influence.

%%%%%%%%%%%%%%%%%%%%%%%%%%%%%%%%%%%%%%%%%%%%%%%%%%%%%%%%%%%%%%%
\section{Recent results}

\begin{figure}
\begin{center}
\includegraphics[width=9cm]{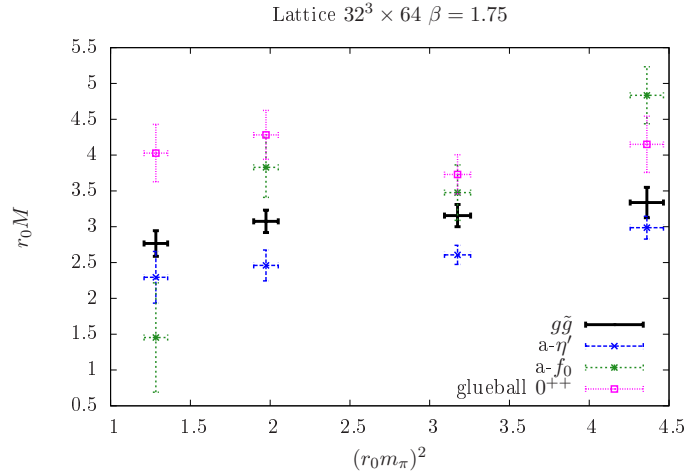} 
\caption{The masses of all particles at $\beta=1.75$ ($a\approx 0.06\
\text{fm}$) obtained on a $32^3\times 64$ lattice ($L=1.8$ fm) with a
statistics of about $5000$ configurations at each point. All masses are
given in units of $r_0$.}
\label{fig:largelat}
\end{center}
\end{figure}

\begin{figure}
\begin{center}
\includegraphics[width=9cm]{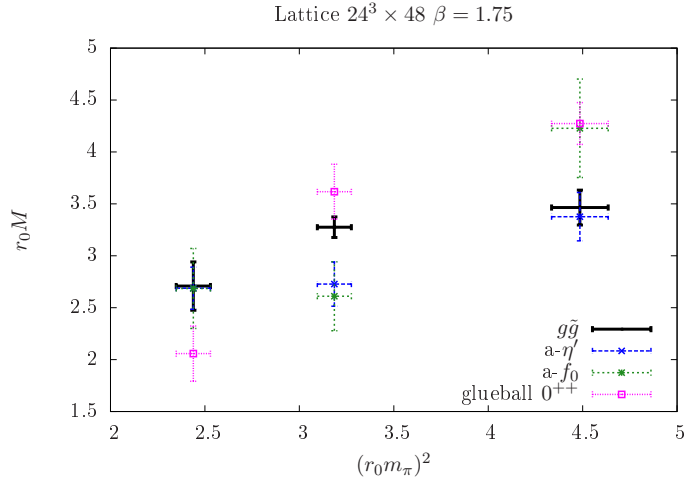} 
\label{fig:smallat}
\caption{The masses of all particles at $\beta=1.75$ ($a\approx 0.06\
\text{fm}$) obtained on a $24^3\times 48$ lattice ($L=1.35$ fm) with a
statistics of about $10,000$ configurations at each point. All masses are
given in units of $r_0$.}
\end{center}
\end{figure}

The results of our investigations of finite size effect and lattice
artifacts are the basis for our current simulations. The influence of the
finite volume seems to be small. A small lattice spacing is important to
keep the influence of the lattice artifacts under control. This result,
obtained from the gluino-glue and the $\aetap$ mass, is already quite
convincing. For a complete investigations of the effects it is, however,
indispensable to get information also form the other particle states. To
distinguish the small mass gap at $\beta=1.75$ from the statistical and
systematic errors, a good signal is needed for all observables. Before we
test further improvements of the action to reduce the lattice artifacts we
have, therefore, made improvements in the measurement of the glueballs and
the scalar mesons. Especially for the glueballs most important for an
improvement is to increase the statistics significantly. Our first
investigations were done at a lattice size of $32^3\times 64$ ($L=1.8$ fm),
see Fig.~\ref{fig:largelat}. In our investigations of the finite size
effects we found that these results are consistent with those at a lattice
size of $24^3\times 48$ ($L=1.35$ fm ).
Based on these findings we have chosen the smaller lattice size and
increased the statistics from about $5000$ configurations at each $\kappa$
to around $10000$. A preliminary summary of these data is shown in
Fig.~\ref{fig:smallat}.

%%%%%%%%%%%%%%%%%%%%%%%%%%%%%%%%%%%%%%%%%%%%%%%%%%%%%%%%%%%%%%%
\section{Conclusions}

We have investigated the influence of the various simulation parameters on
the unexpectedly observed mass gap between the particle states of a
supersymmetric multiplet.

The finite size effects increase the mass gap as found from the splitting of
the $\aetap$ and gluino-glueball mass. The supersymmetry breaking terms that
are introduced by the lattice discretisation seem to get finite size
corrections that enlarge their influence. However, this effect is found to
be small compared to the statistical error at moderate lattice sizes above
around $1.2\ \text{fm}$.

The finite lattice spacing was the most relevant source of supersymmetry
breaking in the previous results of our collaboration. The mass gap at our
current, smaller, lattice spacing is significantly reduced compared to the
previous results. In view of the statistical errors of the observables, it
is already difficult to obtain the mass gap.

To determine the remaining mass gap with enough precision and to answer the
question, whether it persists in the continuum limit, it is necessary to
reduce the statistical errors of the observables. One important parameter
for such an improvement is the statistics. An investigation of the
improvements of the observables is the main task of our current simulations.

Improvements of the action, like an increased level of stout smearing or the
inclusion of a clover term might also help to get a more precise answer
about the relevance of the mass gap in the continuum limit. We have started
to investigate these improvements.

%%%%%%%%%%%%%%%%%%%%%%%%%%%%%%%%%%%%%%%%%%%%%%%%%%%%%%%%%%%%%%%
\section*{Acknowledgments}

This project is supported by the German Science Foundation (DFG) under
contract Mu 757/16, and by the John von Neumann Institute for Computing
(NIC) with grants of computing time. Further computing time has been
provided by the computer cluster PALMA of the University of M\"unster.

%%%%%%%%%%%%%%%%%%%%%%%%%%%%%%%%%%%

\end{document}